%% file: main.tex
\documentclass[12pt]{article}

\usepackage{sbc-template}
\usepackage{graphicx,url}
\usepackage[utf8]{inputenc}
\usepackage[english]{babel}
\usepackage{multirow}
\usepackage{graphicx}
\usepackage{float}
\usepackage{multirow}
\usepackage[nolist]{acronym}
\usepackage{amsmath}

\usepackage{graphicx}
\usepackage{subfig} % For subfigures

\usepackage{tikz}
\newcommand{\circlednumber}[1]{%
    \tikz[baseline=(char.base)]{
        \node[shape=circle, draw=black, fill=black, text=white, inner sep=0.2pt, font=\sffamily\small] (char) {#1};
    }
}
     
\title{Performance Evaluation and Threat Mitigation in Large-scale 5G Core Deployment}

\author{Rodrigo Moreira\inst{1}, Larissa F. {Rodrigues Moreira}\inst{1}, Flávio {de Oliveira Silva}\inst{2}}

\address{Institute of Exact and Technological Sciences -- Federal University of Viçosa
  (UFV)\\
  Rio Paranaíba -- MG -- Brazil
\nextinstitute
  Department of Informatics -- School of Engineering\\
  University of Minho (UMinho) -- Braga -- Portugal
  \email{\{rodrigo, larissa.f.rodrigues\}@ufv.br, flavio@di.uminho.pt,}  
}

\begin{document} 
\input{acronym}

\maketitle

\begin{abstract}
The deployment of large-scale software-based 5G core functions presents significant challenges due to their reliance on optimized and intelligent resource provisioning for their services. Many studies have focused on analyzing the impact of resource allocation for complex deployments using mathematical models, queue theories, or even Artificial Intelligence (AI). This paper elucidates the effects of chaotic workloads, generated by Distributed Denial of Service (DDoS) on different Network Functions (NFs) on User Equipment registration performance. Our findings highlight the necessity of diverse resource profiles to ensure Service-Level Agreement (SLA) compliance in large-scale 5G core deployments. Additionally, our analysis of packet capture approaches demonstrates the potential of kernel-based monitoring for scalable security threat defense. Finally, our empirical evaluation provides insights into the effective deployment of 5G NFs in complex scenarios.
\end{abstract}
     
\section{Introduction}\label{sec:introduction}

The \ac{B5G} network is designed to offer more than connectivity, enabling disruptive applications like human-machine interaction and smart environments~\cite{Tera2025}. \ac{B5G} networks will drive technological transformation, meeting quantitative demands—non-terrestrial networks, higher cell density, \ac{AI}-based optimization, and spectrum sharing—alongside qualitative needs like scalability, sustainability, trust, human-centeredness, and digital inclusion~\cite{Tsekenis2024, Brinton2025}.

Software-based \ac{B5G} cores enable the community to propose and assess both evolutionary and disruptive approaches for the control and data planes within \ac{5G} cores, whether open-source or commercial~\cite{Rouili2024}. Many interventions, such as AI-driven security, focus on specific \ac{NFs}~\cite{Martins2023, Moreira2024, Suomalainen2025}. While existing studies evaluate \ac{5G} cores from various perspectives, the impact of individual \ac{NFs} on user perception and host systems under chaos remains unexplored~\cite{Mukute2024}.

This paper examines the impact of resource constraints on \ac{5G} \ac{NFs} on \ac{UE} perception during registration and \ac{PDU} session establishment. Using a chaos engineering-based method, we apply synthetic workloads to simulate an \ac{DDoS} attached to each \ac{NF}, whereas a \ac{UE} sensor collects registration statistics. We statistically assessed their effects by asynchronously stressing \ac{CPU}, memory, and both. We also analyze the overhead of a container and the \ac{ML}-based threat defense proposed mechanism on the host system, considering user and kernel space sniffing.

The structure of this work is as follows: Section~\ref{sec:related_work} discusses methodologies analogous to the one proposed herein. Section~\ref{sec:evaluation_method} delineates the evaluation methodology. The evaluation results are presented in Section~\ref{sec:experimental_evaluation}, and the study concludes with Section~\ref{sec:concluding_remarks}.

\section{Related Work}\label{sec:related_work}

Research on 5G networks has encompassed diverse aspects, from evaluating simulation and emulation tools to optimizing core network functionalities. \cite{Rouili2024} analyzed open-source 5G RAN tools, comparing \ac{SDR}-based, emulation, and simulation scenarios in terms of throughput, latency, resource use, coverage, and power consumption. 

\cite{Phan2024} developed two software-based 5G Core Network testbeds (one using desktop PCs and another a high-performance server), integrating container orchestration tools to enable scalable and efficient deployment. However, the authors did not consider evaluation metrics like connectivity tests, latency, and packet loss. 
\cite{Bolla2024} addressed the challenge of energy efficiency in virtualized 5G \ac{UPFs} by comparing packetization methods and conducting experiments in diverse traffic scenarios, measuring performance (\ac{CPU} usage and memory usage) and energy consumption.

\cite{Mukute2024} and \cite{Chen2024} investigated performance challenges in open-source 5G Core Networks, particularly free5GC, focusing on control plane functions and scalability, considering metrics such as \ac{CPU} utilization and registration time. \cite{Mukute2024} introduced a benchmarking framework linking system call optimizations to improved user registration performance, while \cite{Chen2024} examined scalability under varying registration scales and duplicate requests.

In contrast to the aforementioned studies, this study advances the state-of-the-art by addressing the deployment challenges of 5G core functions under chaotic workloads such as \ac{DDoS} attacks. By analyzing their impact on \ac{NFs} and \ac{UE} registration performance, as well as demonstrating the potential of kernel-based monitoring for scalable security threat defense and the necessity of tailored resource provisioning for \ac{SLA} compliance.

%\begin{itemize}
  %  \item Evaluating Open-Source 5G SA Testbeds: Unveiling Performance Disparities in RAN Scenarios
   % \item Building a 5G Core Network Testbed: Open-Source Solutions, Lessons Learned, and Research Directions
    %\item Evaluation of Free5GC Forwarding Performance on Private and Public Clouds
    %\item Unsupervised Graph-Sequence Anomaly Detection for 5G Core Network Control Plane Traffic
    %\item Evaluation of UE Registration Performance in Open-Source 5G Core Networks
    %\item Accelerating free5GC Data Plane Using Programmable Hardware
    %\item Design and Implementation of Data Plane Supporting Time-Sensitive Networking in 5G Networks
    %\item Performance Evaluation of 5G Core Network Control-Plane Using Open5GS and Kubernetes
    %\item Control Plane Performance Benchmarking and Feature Analysis of Popular Open-Source 5G Core Networks: OpenAirInterface, Open5GS, and free5GC
    %\item Comparison of the Performance and Energy Efficiency Evaluation of 5G User-Plane Functions
%\end{itemize}

\section{Evaluation Method}\label{sec:evaluation_method}

\begin{figure}[htbp]
  \includegraphics[width=\textwidth]{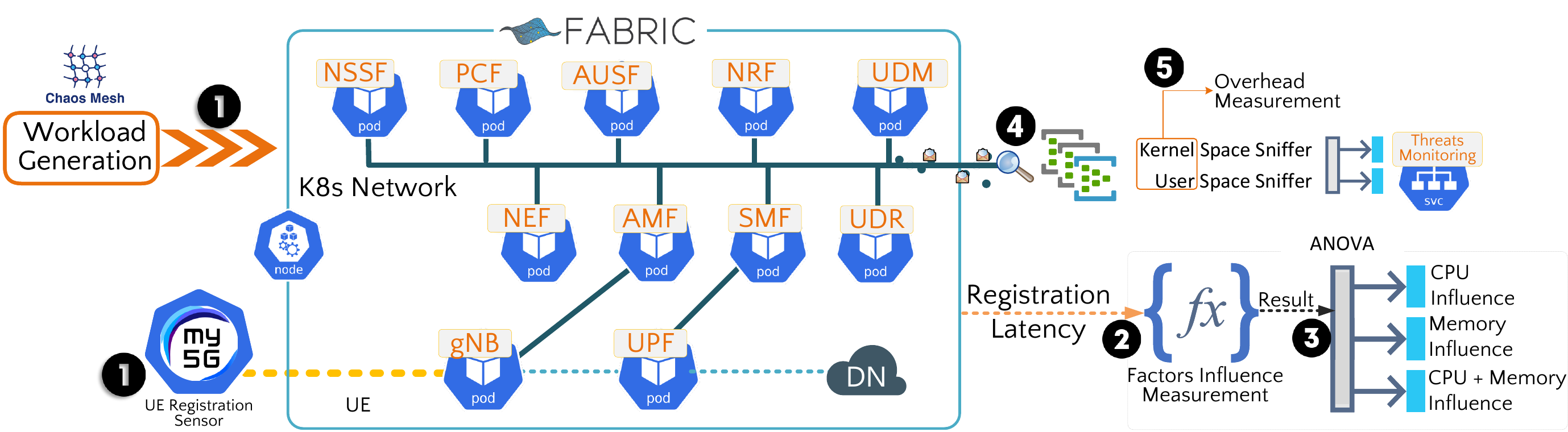}
  \caption{Benchmarking Method.}
  \label{fig:method}
\end{figure}

To evaluate whether \ac{5G} \ac{NFs} influences and under which workload circumstances impact the \ac{UE} registration experience, we propose a benchmarking method based on chaos engineering with factor analysis. Figure~\ref{fig:method} illustrates the proposed method.

Step \circlednumber{1} injects workload parallel to a \ac{UE} application requesting registration and session establishment \ac{PDU} at a constant rate to the core. These requests assume prior \ac{UE} registration in the \ac{5G} core database. The \ac{UE} registration sensor application extends my5G-RANTester~\cite{lucas2022}, with adaptations ensuring a constant duration and rate of requests to extract experimental latencies.

For workload injection into \ac{5G} core components, we used Chaos Mesh, a chaos engineering tool for inducing stress in Kubernetes clusters. Table~\ref{tab:stress_scenarios} details the stressed components, load intensity, and time parameters customizable and empirically defined in our experiments.

\begin{table}[ht]
    \centering
    \caption{Stress Test Scenarios and Parameters.}
    \scriptsize
    \label{tab:stress_scenarios}
    \begin{tabular}{|c|c|c|c|}
        \hline
        \textbf{Scenario} & \textbf{CPU Load (\%)} & \textbf{Memory (MiB)} & \textbf{Duration (s)} \\
        \hline
        CPU Stress & 50 & - & 20 \\
        \hline
        Memory Stress & - & 512 & 20 \\
        \hline
        CPU + Memory Stress & 50 & 512 & 20 \\
        \hline
    \end{tabular}
\end{table}

Step \circlednumber{2} merges sensor application latency data with the timestamps of each load-injection profile in the \ac{5G} core pods. Our deterministic workload induction method records start and end timestamps, enabling correlation between induced load periods and the \ac{UE} registration latency.

Step \circlednumber{3} evaluates the performance impact of induced load on \ac{5G} core components using \ac{ANOVA}. We assess the influence of each \ac{NF} and stress pattern on resources affecting \ac{UE} registration and \ac{PDU} session establishment. Three factors are considered: \ac{CPU}, Memory, and \ac{CPU} + Memory, varied as per Table~\ref{tab:stress_scenarios}. Their effects on \ac{UE} registration time were analyzed using the \ac{LMM} method.

%The \ac{LMM} is a statistical technique that incorporates both fixed and random effects to analyze hierarchical or clustered data. The \ac{LMM} accounts for the fixed effects of \ac{CPU} (\(\beta_1\)), Memory (\(\beta_2\)), and their interaction (\(\beta_3\)), as well as the random effect of different network functions (\(u_j\)). The model is represented as follows:

%\[Y_{ij} = \beta_0 + \beta_1(\text{CPU}_i) + \beta_2(\text{Memory}_i) + \beta_3(\text{CPU} \times \text{Memory}_i) + u_j + \epsilon_{ij}\]

%where \(Y_{ij}\) is the registration time for observation \(i\) in group \(j\), \(\beta_0\) is the intercept, \(u_j\) is the random effect for network functions, and \(\epsilon_{ij}\) is the residual error. This approach allows us to capture the variability and interactions within the system, providing a comprehensive analysis of the factors influencing  \ac{UE} registration latency.

Step \circlednumber{4} employs packet sniffing on the container interface to feed an \ac{ML} model that classifies packets as attacks or normal traffic. Packets are captured using (i) user-space tools (e.g., tcpdump, Wireshark) or (ii) kernel-space methods (\ac{eBPF}). We measure \ac{CPU} and memory usage to assess the overhead of each approach. In Step \circlednumber{5}, discrete statistical methods analyze this impact to determine the most suitable option for production.

\section{Results and Discussion}\label{sec:experimental_evaluation}

We conducted experiments on the Fabric testbed~\cite{baldin2019fabric}, deploying a VM with 16 vCPUs, 32\,GB of RAM, a single-node Kubernetes cluster, Chaos Mesh (a chaos engineering tool), free5GC, and NetData (for monitoring the entire infrastructure). This setup identified the most load-sensitive \ac{5G} core \ac{NF} and its impact on \ac{UE} registration time. Additionally, we evaluated two packet-sniffing methods in containerized environments to assess their monitoring overhead.

For packet sniffing, we used kernel-space (\texttt{ptcpdump}) and user-space (\texttt{Ksniff}) methods, which differ in Kubernetes pod monitoring. \texttt{Ptcpdump} leverages \ac{eBPF} to capture and annotate packets in the kernel. In contrast, \texttt{Ksniff} operates as a \texttt{Kubectl} plugin, uploading a compiled \texttt{tcpdump} binary to the target pod and redirecting its output for analysis.

\subsection{Impact of Attacks on \ac{NF}}\label{subsec:abnormal_traffic_effect}

We first measured how a \ac{DDoS} attack compromises a \ac{5G} \ac{NF} to assess its impact on \ac{UE} registration and \ac{PDU} session establishment. Using concurrent goroutines in Go, we generated high-volume \ac{HTTP} requests to overload the \ac{AMF}, simulating a \ac{DDoS} attack on its endpoint.

\begin{figure}[htbp]
  \centering
  \includegraphics[width=0.4\textwidth]{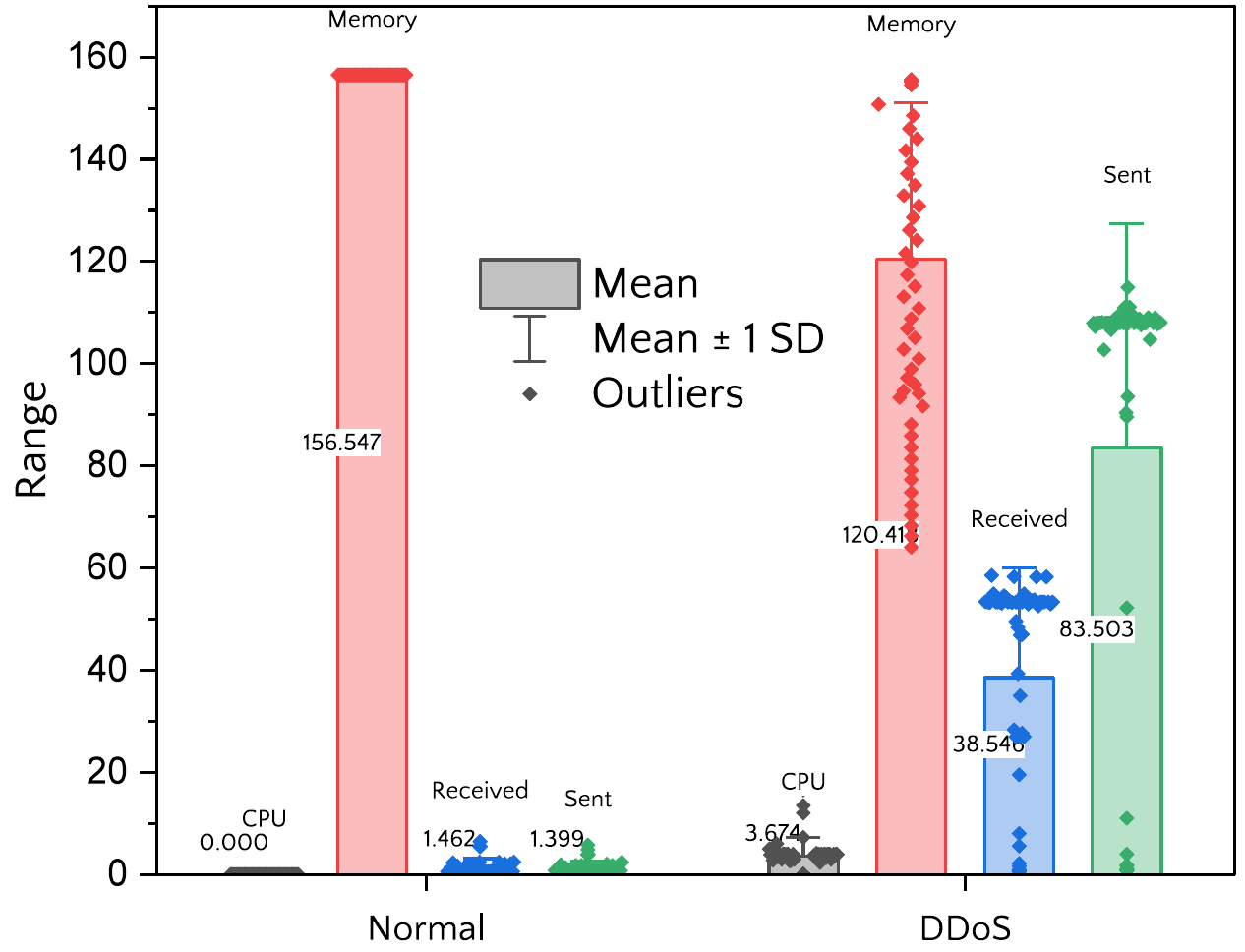}
  \caption{Influence of Attacks on \ac{AMF}.}
  \label{fig:ddos_non-ddos_effect}
\end{figure}

Figure~\ref{fig:ddos_non-ddos_effect} illustrates the impact of a \ac{DDoS} attack on the \ac{AMF}. During the attack, \ac{CPU} utilization spiked to a mean of 3.674, while memory usage dropped to 120.412, indicating disruptions. Network traffic surged, with received and sent data increasing to 38.546 and 83.503, respectively. These results highlight the strain on \ac{AMF} resources, affecting both efficiency and stability, emphasizing the need for advanced detection and mitigation. Additionally, \ac{5G} \ac{NF} workloads can degrade service and impact \ac{UE} quality perception, as discussed in the following subsections.

\subsection{Variability Analyses}\label{subsec:variability_analise}

\begin{figure}[ht]
    \centering
    \begin{tabular}{cc}
			\includegraphics[width=0.425\textwidth]{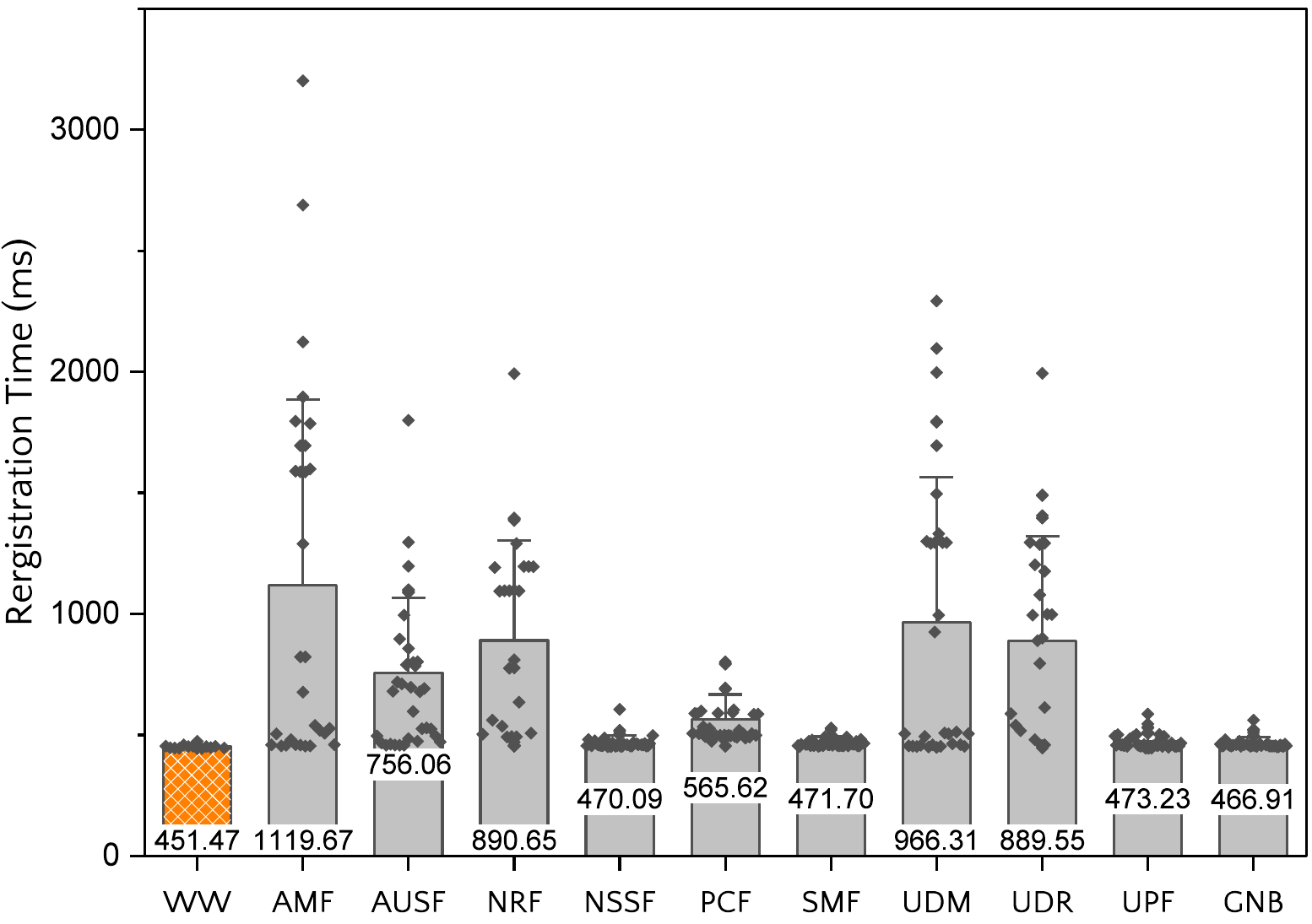} &
			\includegraphics[width=0.4\textwidth]{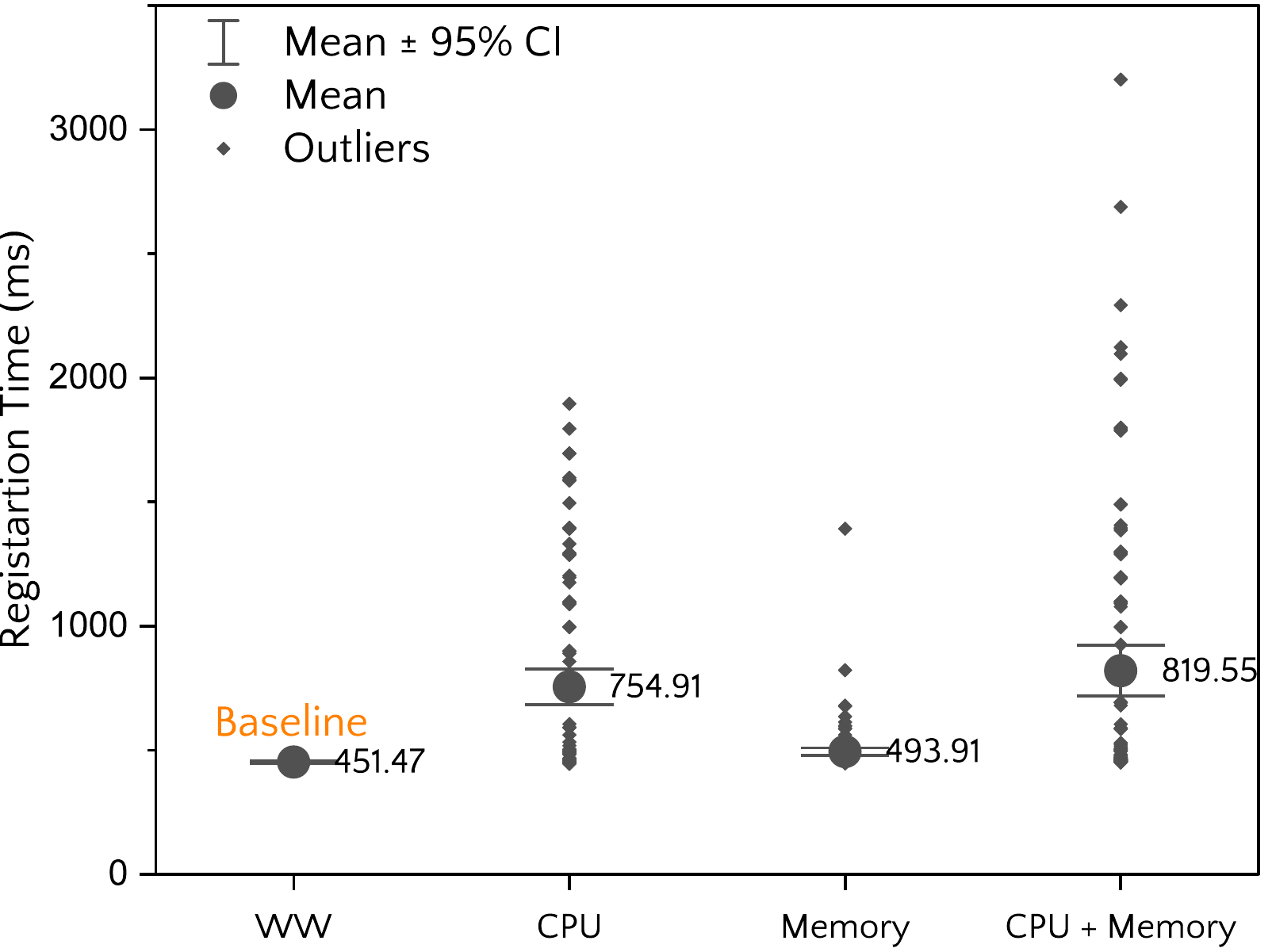} \\
			(a) Registration Time considering different \ac{NFs}. & (b) Induction of different workloads.\\
    \end{tabular}
    \caption{Comparison of variability by \ac{NF}, workloads, and the baseline.}
    \label{fig:variability_comparison}
\end{figure}

Assuming \ac{DDoS} attacks degrade \ac{NFs} performance, we used a workload simulator to assess their impact on \ac{UE} registration. The analysis examined workload-induced variability in open-source \ac{5G} core \ac{NFs}. As shown in Fig.~\ref{fig:variability_comparison}a, \ac{AMF} was the most affected, significantly impacting \ac{UE} registration time and highlighting its critical role under stress.  

As shown in Fig.~\ref{fig:variability_comparison}b, the \ac{CPU} + Memory workload has the higher impact on \ac{UE} registration time in the open-source \ac{5G} core. Thus, in microservices-based \ac{5G} deployments, varying \ac{CPU} and Memory demands across \ac{NFs} must be considered for production environments.

After identifying the variability of the registration time considering the \ac{NF} and the type of workload, we sought to determine whether the \ac{NF} influences the response time as a random effect or whether different \ac{NFs} can affect the registration time, thus being a co-variate in the analysis. To make this decision, a test was conducted to verify whether there was a significant difference between the average times for \ac{NFs}.

We conducted an \ac{ANOVA} test to assess the impact of \ac{NF} on Registration Time. The results (Table~\ref{tab:anova_for_NF_in_registration_time}) show that \ac{NF} is highly significant (\textit{p} $<$ 0.001), with an F-statistic of 18.611059. This confirms that \ac{NF} workload influences \ac{UE} registration as a random effect.

\begin{table}[ht]
\centering
\caption{ANOVA Results for \ac{NF} Influence on Registration Time.}
\resizebox{\textwidth}{!}{%
\begin{tabular}{lcccc}
\hline
Source    & Sum of Squares (\( SS \))  & Degrees of Freedom (\( df \)) & F-Statistic (\( F \)) & \textit{p}-Value (\( p \)) \\
\hline
C(NF)     & \( 1.972265 \times 10^7 \) & 9                            & 18.611059             & \( 1.352433 \times 10^{-25} \) \\
Residual  & \( 4.356657 \times 10^7 \) & 370                          & -                     & -                 \\
\hline
\end{tabular}
}
\label{tab:anova_for_NF_in_registration_time}
\end{table}

%The sum of squares for the \ac{NF} is \( SS_{nf} = 1.972265 \times 10^7 \) with 9 degrees of freedom, and the residual sum of squares is \( SS_{res} = 4.356657 \times 10^7 \) with 370 degrees of freedom. The F-statistic for \ac{NF} is calculated as follows:

%\[ F = \frac{MS_{nf}}{MS_{res}} \]

%Where \( MS_{nf} \) is the mean square for the factor (NF), calculated as \( \frac{SS_{nf}}{df_{nf}} \). \( MS_{res} \) is the mean square for the residuals, calculated as \( \frac{SS_{res}}{df_{res}} \). The terms \( SS \) and \( df \) represent the sum of squares and degrees of freedom, respectively. The full formula is:

%\[ F = \frac{\frac{SS_{nf}}{df_{nf}}}{\frac{SS_{res}}{df_{res}}} \]

%Substituting the values obtained in your analysis: \( SS_{nf} = 1.972265 \times 10^7 \), \( df_{nf} = 9 \), \( SS_{res} = 4.356657 \times 10^7 \), \( df_{res} = 370 \) we get:

%\[ F = \frac{\frac{1.972265 \times 10^7}{9}}{\frac{4.356657 \times 10^7}{370}} \approx 18.611059 \]

%This means that the resulting F-statistic is approximately 18.611059, which indicates the ratio of the variability explained by the factors compared to the residual variability. A high F-value suggests that the factor \ac{NF} has a significant influence on the Registration Time. This justifies the inclusion of \ac{NF} as a variable in the model to account for variations between different network functions.

\subsection{Influence Analyses}\label{subsec:influence_analyze}

The third part of our experiment examines the effect of each \ac{NF} on registration time using a three-way \ac{ANOVA}, with \ac{NF} as a random effect in the \ac{LMM}. We analyzed CPU, Memory, and CPU+Memory workloads to assess their impact. With \ac{NF} showing significance in prior analysis, we included \ac{LMM} as a random effect. The baseline registration time without workload injection was 451.46\,ms (standard deviation: 7.80\,ms). The \ac{LMM} results are summarized below:

%The \ac{LMM} was specified as follows:

%\[
%Y_{ij} = \beta_0 + \beta_1 (\text{CPU}_i) + \beta_2 (\text{Memory}_i) + \beta_3 (\text{CPU} \times \text{Memory}_i) + u_j + \epsilon_{ij}
%\]

%where \(Y_{ij}\) is the Registration Time for observation \(i\) in group \(j\), \(\beta_0\) is the intercept (baseline), \(\beta_1, \beta_2, \beta_3\) are the fixed effects coefficients for CPU, Memory, and their interaction, \(u_j\) represents the random effect for NF, and \(\epsilon_{ij}\) is the residual error.

%The results from the \ac{LMM} are summarized below:

\begin{table}[ht]
\centering
\scriptsize
\caption{Mixed Linear Model Regression Results.}
%\resizebox{\textwidth}{!}{%
\begin{tabular}{lccccc}
\hline
Source & Coefficient ($\beta$) & Std. Error & z-value & \textit{p}-value & 95\% CI \\
\hline
Intercept & 451.467 & 270.491 & 1.669 & 0.095 & [-78.685, 981.619] \\
C(stress\_test)[T.CPU] & 352.239 & 283.976 & 1.240 & 0.215 & [-204.344, 908.821] \\
C(stress\_test)[T.Memory] & 47.211 & 283.729 & 0.166 & 0.868 & [-508.886, 603.309] \\
C(stress\_test)[T.CPU$+$Memory] & 420.092 & 284.012 & 1.479 & 0.139 & [-136.561, 976.745] \\
Group Var & 67402.169 & 113.331 & - & - \\
\hline
\end{tabular}
%}
\label{tab:a}
\end{table}

%The intercept (baseline) represents the average registration time with no workload injected, which is 451.467 ms. The coefficient for the CPU factor alone (C(stress\_test)[T.CPU]) is 352.239 ms, but it is not statistically significant (\( p = 0.215 \)). The coefficient for the Memory factor alone (C(stress\_test)[T.Memory]) is 47.211 ms, and it is also not statistically significant (\( p = 0.868 \)). The coefficient for the combined CPU and Memory factor (C(stress\_test)[T.CPU$+$Memory]) is 420.092 ms, which is not statistically significant either (\( p = 0.139 \)). The Group Variance was estimated at 67402.169, suggesting considerable variability among the \ac{NFs}.

%The results from the \ac{LMM} indicate that none of the stress test conditions (CPU, Memory, and their combination) exert a statistically significant influence on the registration time. However, among the tested conditions, the combined CPU and Memory stress condition (\texttt{C(stress\_test)[T.CPU$+$Memory]}) has the largest coefficient (\( 420.092 \) ms) with the lowest \textit{p}-value (\( p = 0.139 \)), suggesting it has the most substantial, albeit not statistically significant, effect on the registration time. The variance associated with NF (\( \hat{\sigma}^2_u = 67402.169 \)) contributes considerably to the total variability in registration time, highlighting the importance of accounting for NF as a random effect.

The intercept represents the average registration time without workload injection (\( 451.467 \) ms). The CPU stress test increases registration time by \( 352.239 \) ms (\( p = 0.215 \)), memory stress by \( 47.211 \) ms (\( p = 0.868 \)), and their combination by \( 420.092 \) ms (\( p = 0.139 \)), but none are statistically significant. The group variance (\( 67402.169 \)) indicates substantial variability among the \ac{NFs}.  

The \ac{LMM} results show that while stress conditions do not significantly impact registration time, the combined CPU and memory stress condition (\texttt{C(stress\_test)[T.CPU$+$Memory]}) has the largest effect. The variance associated with \ac{NF} (\( \hat{\sigma}^2_u = 67402.169 \)) contributes considerably to the total variability, highlighting the importance of accounting for \ac{NF} as a random effect.

\begin{figure}[htbp]
  \centering
  \includegraphics[width=0.4\textwidth]{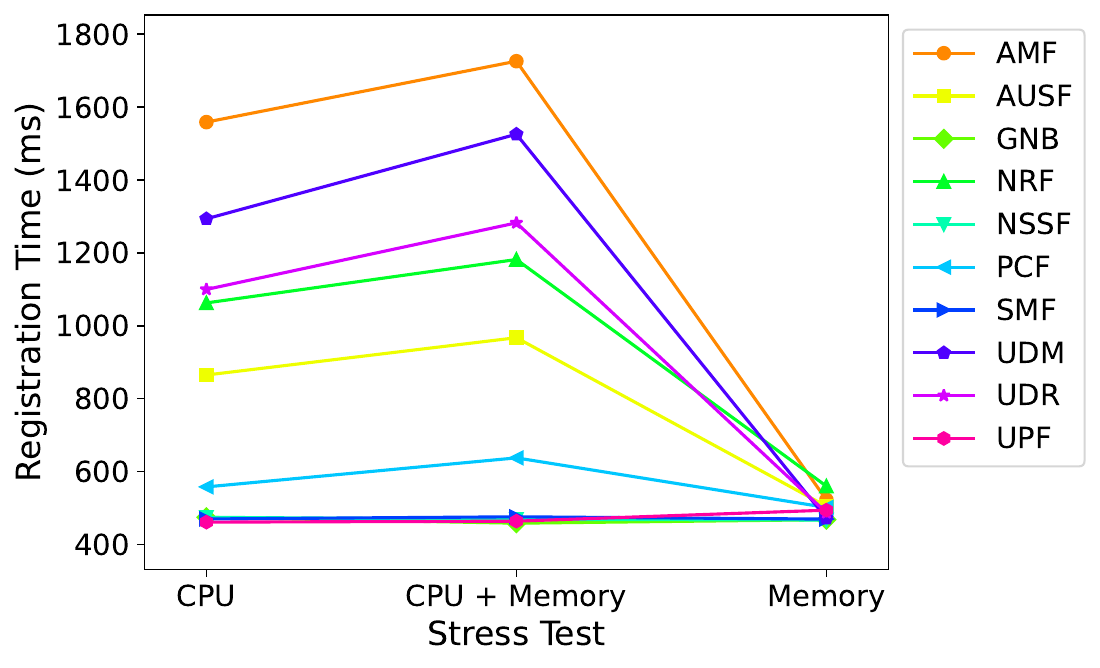}
  \caption{Workload interaction on \ac{NFs}.}
  \label{fig:influence_effects_rank_by_nf}
\end{figure}

From Figure~\ref{fig:influence_effects_rank_by_nf}, we observe that \ac{AMF} has the greatest influence on the registration time, followed by \ac{UDM} and \ac{UDR}. This influence is more pronounced when the \ac{CPU} and memory factors are combined, implying greater difficulty in handling the \ac{UE} registration workload. For all \ac{NFs}, memory stress alone does not have a relevant impact on \ac{UE} registration. Therefore, in large-scale deployments, it is crucial to consider the differences in resource demands of each \ac{NF}.

%Therefore, we can observe from Figure~\ref{fig:influence_effects_rank_by_nf} that \ac{AMF} has the greatest influence on the Registration Time, followed by \ac{UDM} and \ac{UDR}. This influence is more pronounced when the \ac{CPU} and memory factors are combined, which implies a greater difficulty in dealing with the \ac{UE} registration workload. For all \ac{NFs}, the induction of chaos in the memory does not have a relevant influence on \ac{UE} registration. Therefore, in large-scale deployment, it is important to consider the difference in the resource demands of each \ac{NF}.

\subsection{Monitoring Overhead}

To deploy a threat defense mechanism against service degradation perceived by the \ac{UE}, we propose monitoring packet exchanges between \ac{NFs} by sniffing packets and transmitting their hexadecimal representation to an \ac{ML}-based service for classification as normal or \ac{DDoS}. This method introduces additional workload to the cluster, which we analyze in detail.

\begin{figure}[htbp]
    \centering
    \begin{tabular}{cc}
			\includegraphics[width=0.46\textwidth]{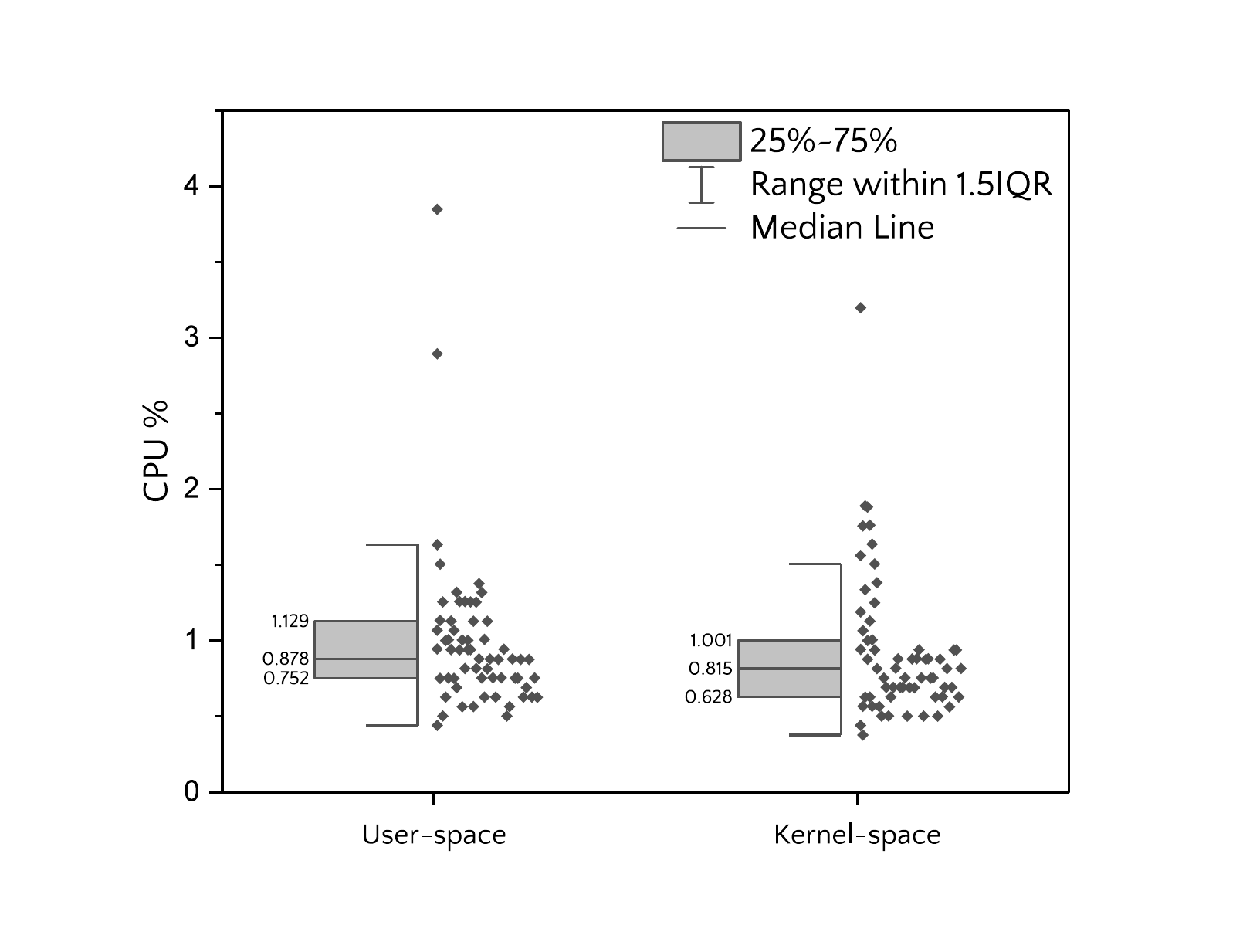} &
			\includegraphics[width=0.46\textwidth]{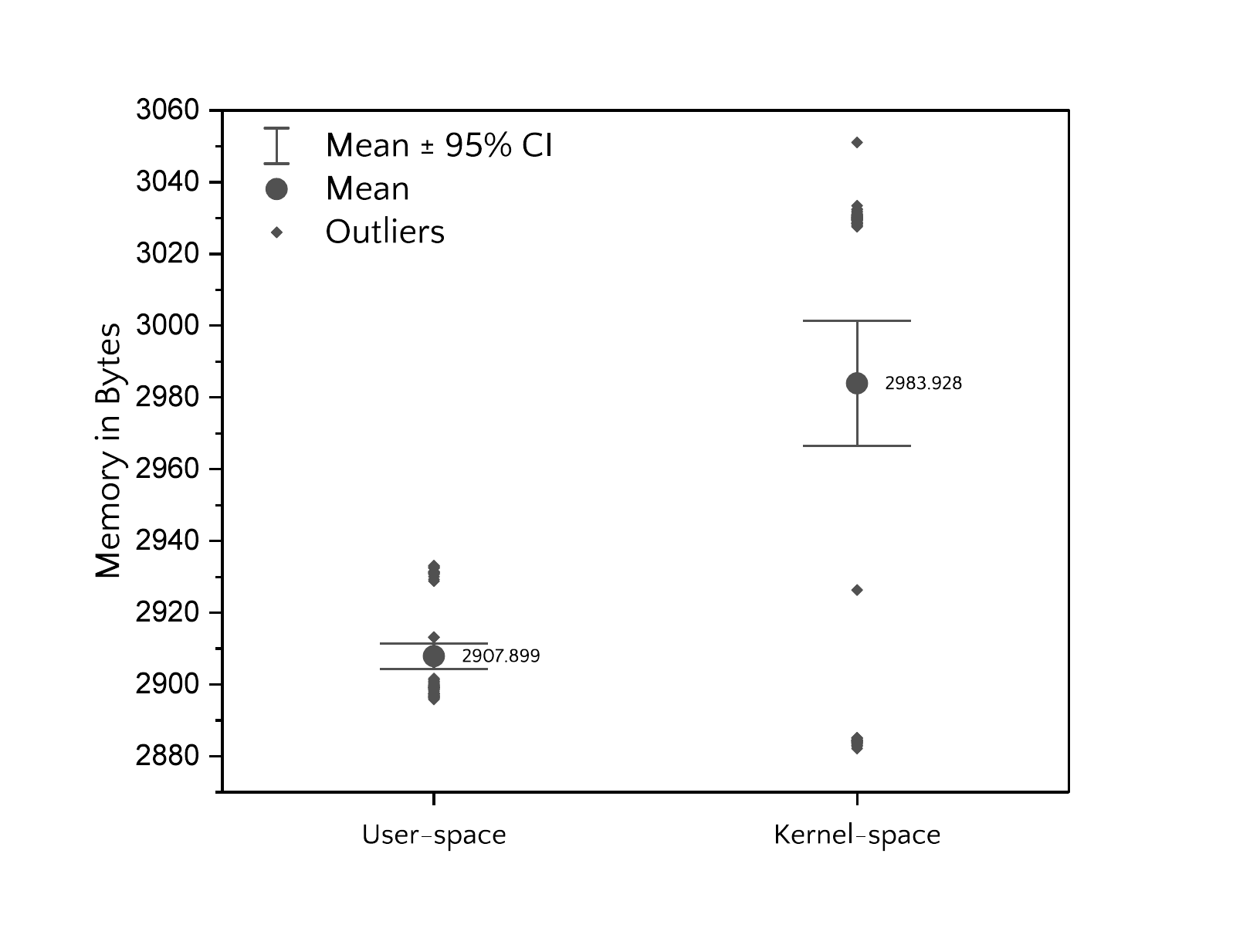} \\
			(a) \ac{CPU} & (b) Memory.\\
    \end{tabular}
    \caption{Comparison of packet sniffing methods.}
    \label{fig:memory_and_cpu_overhead}
\end{figure}

The analysis of \ac{CPU} consumption (Figure~\ref{fig:memory_and_cpu_overhead}a) revealed that the kernel-based method using \ac{eBPF} had slightly lower \ac{CPU} usage than the user-space method with Ksniff. While the median \ac{CPU} consumption was similar (0.878\% for user-space and 0.815\% for kernel), the kernel-based approach showed greater amplitude in the third quartile and less data dispersion, indicating \ac{eBPF} efficiently captures packets within the kernel.

The memory usage analysis (Figure~\ref{fig:memory_and_cpu_overhead}b) indicates that the kernel-based method has a higher mean memory consumption (2983.928 bytes) compared to the user-space approach (2907.899 bytes), implying greater memory overhead. Additionally, both methods exhibit outliers, with higher dispersion in kernel-space data, indicating variability in memory performance. These sniffed packets could support an ML-based security threat defense, leveraging the kernel-space approach to enhance security in such deployments.  

Using a single virtual machine simplifies deployment but limits the realism of a distributed 5G core. Similarly, synthetic traffic lacks real-world variability, which future work aims to address.

\section{Concluding Remarks}\label{sec:concluding_remarks}

This paper examines the impact of varying loads on \ac{5G} control plane \ac{NFs} and their effect on user perception. While prior methodologies assessed different \ac{5G} open source cores, they did not quantify the latency impact of each \ac{NF} on \ac{UE} registration and \ac{PDU} session establishment. This research offers quantifications to inform large-scale deployments, enabling differential resource allocation for \ac{NFs} in resource-constrained environments, particularly under low-power policies.

The findings indicate that the \ac{AMF} entity requires the most resources due to its significant impact on \ac{UE} registration time. Kernel-based monitoring approaches like \ac{eBPF} can improve security threat detection while maintaining efficiency in resource-constrained 5G deployments. Future research will assess failure resilience in open-source \ac{5G} cores, providing a reference for large-scale deployments that meet the resource needs of each \ac{NF}.

\section*{Acknowledgments}

%We acknowledge the financial support of the FAPESP MCTIC/CGI Research project 2018/23097-3 and FAPEMIG (Grant APQ00923-24). We also thank the National Council for Scientific and Technological Development (CNPq) under grant number 421944/2021-8 (call CNPq/MCTI/FNDCT 18/2021) and Centro ALGORITMI, funded by Fundação para a Ciência e Tecnologia (FCT) within the RD Units Project Scope 2020-2023 (UIDB/00319/2020) for partially support this work.

We acknowledge the financial support of the FAPESP MCTIC/CGI Research project 2018/23097-3 and FAPEMIG (Grant APQ00923-24). We also thank FCT—Fundação para a Ciência e Tecnologia within the R\&D Unit Project Scope UID/00319/Centro ALGORITMI (ALGORITMI/UM) for partially supporting this work.

\bibliographystyle{sbc}
\bibliography{sbc-template}

\end{document}

%% file: acronym.tex
%Numbering
\acrodef{3GPP}{3rd Generation Partnership Project}
%-----A-----
\acrodef{AI}{Artificial Intelligence}
\acrodef{AMF}{Access and Mobility Management Function}
\acrodef{ANOVA}{Analysis of Variance}
%-----B-----
\acrodef{B5G}{Beyond Fifth Generation}
\acrodef{BPF}{Berkeley Packet Filter}
\acrodef{eBPF}{Extended Berkeley Packet Filter}
%-----C-----
\acrodef{CPU}{Central Processing Unit}
%-----D-----
\acrodef{DoS}{Denial of Service}
\acrodef{DDoS}{Distributed Denial of Service}
\acrodef{DNN}{Deep Neural Network}
\acrodef{DRL}{Deep Reinforcement Learning}
\acrodef{DT}{Decision Tree}
%-----E-----
\acrodef{ETSI}{European Telecommunications Standards Institute}
%-----F-----
\acrodef{FIBRE}{Future Internet Brazilian Environment for Experimentation}
\acrodef{FIBRE-NG}{Future Internet Brazilian Environment for Experimentation New Generation}
\acrodef{5G}{Fifth-generation of Mobile Telecommunications Technology}
%-----G-----
\acrodef{GNN}{Graph Neural Networks}
%-----H-----
\acrodef{HTM}{Hierarchical Temporal Memory}
\acrodef{HTTP}{Hypertext Transfer Protocol}

%-----I-----
\acrodef{IAM}{Identity And Access Management}
\acrodef{IID}{Informally, Identically Distributed}
\acrodef{IoE}{Internet of Everything}
\acrodef{IoT}{Internet of Things}
%-----J-----
%-----K-----
\acrodef{KNN}{K-Nearest Neighbors}
%-----L-----
\acrodef{LSTM}{Long Short-Term Memory}
\acrodef{LMM}{Linear Mixed Model}
%-----M-----
\acrodef{M2M}{Machine to Machine}
\acrodef{MAE}{Mean Absolute Error}
\acrodef{ML}{Machine Learning}
\acrodef{MOS}{Mean Opinion Score}
\acrodef{MAPE}{Mean Absolute Percentage Error}
\acrodef{MSE}{Mean Squared Error}
\acrodef{mMTC}{Massive Machine Type Communications}
\acrodef{MFA}{Multi-factor Authentication}
\acrodef{MQTT}{Message Queuing Telemetry Transport}

%-----N-----
\acrodef{NF}{Network Function}
\acrodef{NFs}{Network Functions}
%-----O-----
\acrodef{OSM}{Open Source MANO}
%-----P-----
\acrodef{PDU}{Protocol Data Unit}
%-----Q-----
\acrodef{QoE}{Quality of experience}
\acrodef{QoS}{Quality of Service}
%-----R-----
\acrodef{RAM}{Random-Access Memory}
\acrodef{RF}{Random Forest}
\acrodef{RL}{Reinforcement Learning}
\acrodef{RMSE}{Root Mean Square Error}
\acrodef{RNN}{Recurrent Neural Network}
\acrodef{RAM}{Random-Access Memory}

%-----S-----
\acrodef{SDN}{Software-Defined Networking}
\acrodef{SFI2}{Slicing Future Internet Infrastructures}
\acrodef{SLA}{Service-Level Agreement}
\acrodef{SON}{Self-Organizing Network}
\acrodef{SDR}{Software Defined Radios}
\acrodef{SCTP}{Stream Control Transmission Protocol}

%-----T-----
%-----U-----
\acrodef{UE}{User Equipment}
\acrodef{UEs}{User Equipments}
\acrodef{UDM}{Unified Data Management}
\acrodef{UDR}{Unified Data Repository}
\acrodef{UPFs}{User-Plane Functions}
%-----V-----
\acrodef{VoD}{Video on Demand}
\acrodef{VR}{Virtual Reality}
\acrodef{V2X}{Vehicle-to-Everything}

%-----W-----
%-----X-----
%-----Y-----
%-----Z-----